# Implementation of a cathode directed streamer model in Air under different voltage stresses.

F. Boakye-Mensah[1]* , N. Bonifaci[1], R. Hanna[1], I. Niyonzima[1]

[1]Univ. Grenoble Alpes, CNRS, Grenoble INP*, G2Elab, Grenoble, 38031, France
*francis.boakye-mensah@g2elab.grenoble-inp.fr

*Abstract* - To find a viable alternative to $SF_6$ with growing climate change regulations, proper evaluation of alternatives such as compressed air ought to be done. For medium voltage applications, the withstand voltage is used as the dimensioning criteria and this is dependent on the initiation and propagation of streamers which are precursors to electrical breakdown. For design optimization, a thorough understanding of the initiation and propagation mechanisms of such electrical discharges under different stresses, pressure etc. ought to be studied experimentally and numerically also via a predictive model. Most of the numerical studies have so far been done via homemade codes as streamer models are not readily available in commercial software because of the complexity and non-linearity of such computations. Recently, with the increased robustness of the plasma module of the commercial finite element software, COMSOL™ Multiphysics, streamer discharge models can be developed with reasonable accuracy.

In this paper, an implementation and validation approach is presented for streamer evolution in air for different voltage stresses. Results of simulations for short gaps (≤ 5 mm) under Standard Temperature and Pressure (STP) conditions have been presented, analyzed and compared with some classical papers to evaluate the suitability of such a model for further studies of non-thermal electrical discharges.

## 1. INTRODUCTION

Sulphur Hexafluoride, $SF_6$ is widely used as an insulation gas for medium and high voltage electrical equipment. Notwithstanding its excellent physical properties (thermal and chemical stability, low boiling point, high dielectric strength etc.), it is characterized by a strong greenhouse effect. With increasing environmental regulations, it is imperative to find eco-friendly gas alternatives. For medium voltage applications, compressed air and other solid-gas combinations are viable alternatives but the withstand voltage of compressed air is approximately a third of that of $SF_6$ and thus requires a higher pressure – distance product for the same application in accordance with Paschen's law. To improve the withstand and test voltages with dielectric introduction, a thorough understanding of the electrical discharges that occur in the gas is required. A starting point for this study are streamer discharges which initiate in locations with sharp and curved edges of conductors exposed to sufficiently high voltages. With this enhancement of the local electric field, the streamer can propagate for a significant distance in the medium and depending on the gap distance, eventually bridge it. This propagation is supported by strong electric fields at the propagating plasma fronts.

In electrical equipment, the bridging of the gap may or may not lead to breakdown. Knowledge of electrical breakdown is nonetheless essential for equipment design. Computer models present analytical and predictive tools for streamer discharges however the numerical solution of streamer problems are highly nonlinear and computationally expensive and require special numerical techniques. Most of recent research on streamer dynamics have been conducted using homemade finite element [1] – [6] and finite volume codes [7] – [10] but for equipment manufacturers, the duplication of such works in commercial software is essential because of the user friendly environment and that is what has been achieved using the plasma module of COMSOL™ Multiphysics in this paper.

## 2. GOVERNING EQUATIONS

A hydrodynamic (drift-diffusion) approach is used for the formulation of the streamer propagation model within which variations of densities of electrons, positive ions (cations) and negative ions (anions) in space and time are considered ( [5], [11]). The model describes the generation, losses and the motion of the three charged species, and the formulation results in Partial Differential Equations (PDEs) that account for the rates of the physical processes aforementioned. This can be represented logarithmically by (1) and (2) below [6]:

$$\frac{\partial n_{xl}}{\partial t} + \nabla(-D_x \nabla. n_{xl} - \mu_x n_{xl} \vec{E}) = R_{xl} \qquad (1)$$

$$n_{xl} = \ln(n_e, n_p, n_n) \qquad (2)$$

$n_e$, $n_p$ and $n_n$ denote the densities of electrons, cations and anions respectively, $m^{-3}$; µ is the mobility of the charged species, $m^2/Vs$; $D$ is the diffusion coefficient, $m^2/s$; $R$ is the net rate of the generation and loss processes; $m^{-3}s^{-1}$; $\vec{E}$ is the electric field, $V/m$; and $t$ is for time. Generally, the mobility of positive and negative ions are two orders of magnitude lower than the mobility of electrons and thus for short streamers (streamers over relatively small gap distances less than 5 mm), drift of ions can be neglected. This is because the time scale of the streamer process is in nano-seconds. The logarithmic implementation of the charge species is done to prevent negative value concentrations and thus improve simulation stability.

The rate processes ($R$) can be grouped into generation terms and loss terms. The generation terms include photoionization $S_{ph}$; effective ionization, $S_{eff} = (\alpha - \eta)\, n_{el}\, \mu_e\, \vec{E}$ and background ionization $S_b$ while the loss terms are electron - ion recombination, $L_{ep} = \beta_{ep}\, n_{el}\, n_{pl}$; and ion - ion recombination, $L_{pn} = \beta_{pn}\, n_{pl}\, n_{nl}$. In the preceding expressions, $\alpha$ is the Townsend's ionization coefficient, $m^{-1}$; $\eta$ is the attachment coefficient, $m^{-1}$; $\beta$ is the respective recombination coefficient, $m^3/s$.

The electric field is determined using the Poisson's equation for electric potential $\phi$. The solution provides electric field distributions affected by the space charge which is used to compute the kinetic coefficients and the rates of individual processes [5]:

$$\nabla(\varepsilon_0 \varepsilon_r \nabla \phi) = -e(n_p - n_e - n_n) \quad (3)$$
$$\vec{E} = -\nabla \phi \quad (4)$$

Here, $e$ is the elementary charge of an electron, $\varepsilon_0$ represents the permittivity of vacuum and $\varepsilon_r$ is the dielectric constant of the gas.

The partial differential equations together with the Poisson's equation for electric potential with proper initial values and boundary conditions as well as well-defined kinetic and rate processes form a self-consistent model.

## 3. MODEL IMPLEMENTATION

The model has been implemented in COMSOL™ Multiphysics, a commercial finite element solver using the plasma module (plas). Documentation on the plasma module can be found in [12], [13].
Standard Temperature and Pressure (STP) are being used here and thus the density $N$, can be computed using the equation:

$$N = \frac{P}{kT} \quad (5)$$

Where $k$ is the Boltzmann's constant.

Since the plasma chemistry was used, a set of collision cross sections for air were specified to represent the pertinent physical processes that are necessary for streamer modelling like ionization, attachment etc. and also for the proper functioning of the model (boundary conditions). For the cross sectional species and swarm parameters used, refer to appendix A. The parameters were introduced into the model as interpolation tables dependent on the reduced electric field $E/N$.

To reduce the computational costs, a 2D axisymmetric model for a point to plane geometry using a needle for the anode and a grounded metal plate for the cathode is adopted. This has been selected so that at least one of the electrodes form a region with a strong electrostatic field.

Propagating streamers require fine computational mesh to accurately resolve the charged layers in the vicinity of the electrodes and in areas with high gradients of charge density, most especially, the streamer head. An Adaptive Mesh Refinement (AMR) is adopted for this purpose. This allows for the implementation of a coarse mesh throughout the computational domain therefore reducing the computational time. To be able to run simulations for larger gap spaces and computational domains, partitions have been employed in the computational domain so as to be able to utilize different mesh sizes depending on activity levels. In a domain of 1 - 1.5 mm in the r direction depending on the size of the computational domain, a fixed cell grid of 8 μm is utilized and beyond this region, the grid expands according to geometric progression. It is also worth noting that mesh size of 2 μm is used to follow the propagation of the streamer.

For solver settings, a fully coupled nonlinear node with a direct solver is employed and for time stepping, a BDF formula of order 1 – 5 has been used.

## 4. SIMULATION RESULTS AND DISCUSSION

Comparing positive streamers to negative ones, the discharge initiation occurs more easily at relatively lower voltages [6], and this makes it more critical to electrical systems design and thus the evolution of the positive streamer has been studied.

*A. Validation*

To validate the use of the COMSOL™ plasma module for streamer simulations, a comparison of the results obtained using our implementation has been done with the one obtained by Ducasse et al in [4] for the streamer in Air under similar conditions. Recapping, a point to plane geometry with computational domain boundaries of 2 x 2 mm and needle to plane electrode spacing of 1 mm is used. A voltage of +3 kV is applied to the needle and the plane was grounded. A direct comparison between the two works is provided in table 1.

Our implementation mostly varied from the initial work in two ways:
1. No Gaussian seed was used for the initiation of the streamer.
2. The background ionization term $S_b$ used was an order of magnitude lower than the original work. This was done so as to accurately compare the results at various axial positions and time steps.

In both works, the Laplacian electric field was 600 Td.

*Table 1: Comparison between current work and original work by Ducasse et al [4].*

|  | Current work | Ducasse et al [4] |
|---|---|---|
| Implementation | Plasma fluid model + Poisson's Equation | Classical fluid model + Poisson's Equation |
| Meshing Technique | Partitioned domain + AMR min 2 μm, max 8 μm for r < 1 mm above which larger grid size. | Partitioned domain dz = 2.5 μm, dr = 2 μm over 200 μm above which larger grid size. |
| Initial Seed | No initial seed | $N_0 = 10^{14} m^{-3}$ $\sigma_0 = 25$ μm $z_0 = 800$ μm |
| $S_b$ $[1/(m^3.s)]$ | $3 \times 10^{25}$ | $10^{26}$ |

Figures 1 and 2 below show the variation of the electron density profile and electric field distribution along the axial line as the streamer propagates at times 1 ns, 2 ns and 3 ns.

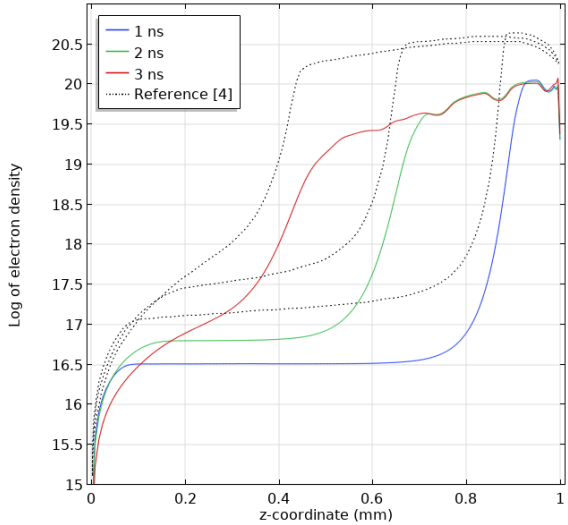

*Figure 1: Electron density variation on streamer axis at times t = 1, 2 and 3 ns for our model (solid line) and original work [4] (dotted line).*

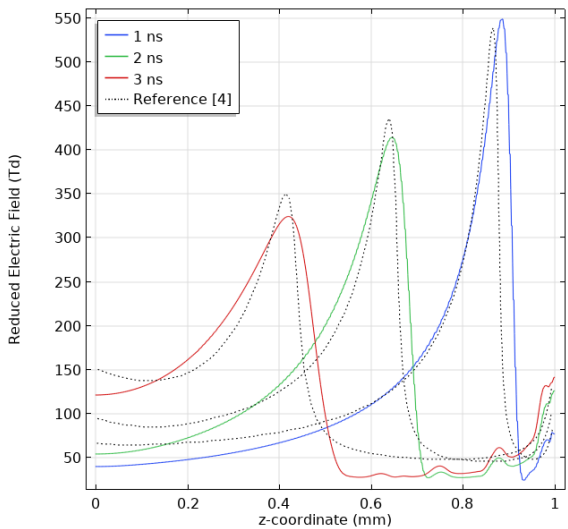

*Figure 2: Reduced electric-field variation on the streamer axis corresponding to figure 1.*

Generally, the two models are in agreeance although small variations can be noticed in the axial displacement of the streamer in both works and these can be attributed to the grid refinement technique and the treatment of the background ionization term in current work (An order of magnitude lower; refer to table 1). Specifically, the electron charge density in the head of the streamer for the current work ranges between $5 - 9 \times 10^{19} m^{-3}$ in contrast to $1 - 5 \times 10^{20} m^{-3}$ in [4].

The streamer crosses the 1 mm gap in 4.9 ns in current work as compared to 5 ns in the original work. In both models, the radius of the streamer at stable propagation was 0.2 mm with a streamer velocity of $\sim 0.2 \ mm/ns$.

### B. Streamer in Short Gaps (5 mm)

Upon verification, positive streamer simulations have been done for a relatively larger gap spacing of 5 mm previously attempted in COMSOL in [5] [6]. The streamer initiates from the background ionization $S_b = 10^{23} \ 1/(m^3.s)$ used as an alternate for a photoionization model. A positive voltage of +15 kV is applied to the needle electrode.

The distribution of electron densities along the symmetry axis has been presented in figure 3. It can be observed that two main stages are evident for the streamer initiation and propagation mechanism; the initiation and radial expansion phase and the propagation phase. Also, from the electric field diagrams in figure 4, one can observe the enhanced field at the tip of the streamer in comparison to the drastically reduced field strength in the channel. The streamer eventually crosses the 5 mm gap in 4.1 ns thus the estimated average front velocity is $\sim 1.2 \ mm/ns$ which is in agreement with the earlier works [5], [6] and falls between $10^5 \ and \ 10^7 \ m/s$ propagating velocity consistent in literature for short streamers [14]. The channel radius at stable propagation computed from the electron reaction rate is also in agreement with earlier works falling between 400 and 600 μm.

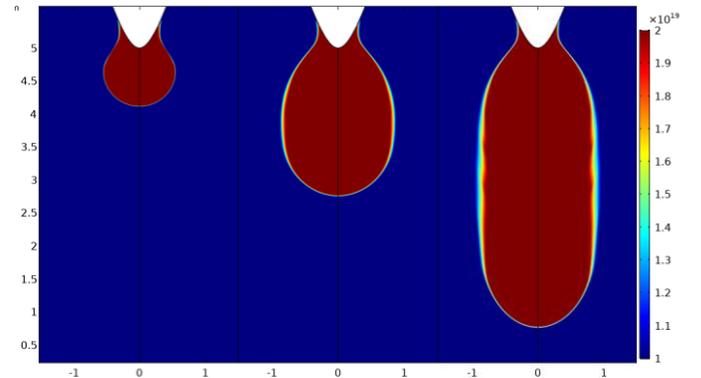

*Figure 3: 2D surface plots of electron density profiles of a streamer propagating in a 5 mm air gap at times t = 1.0, 2.5 and 3.9 ns.*

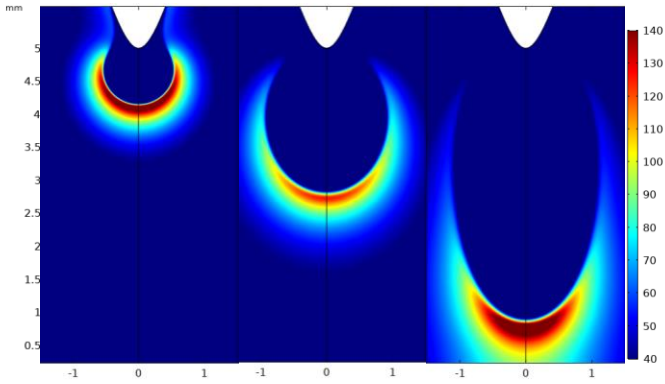

*Figure 4: 2D surface plots of electric field distributions corresponding to electron densities in figure 3.*

From the line diagram for the electron concentration in figure 5, the absolute concentrations in logarithmic scale are presented which corresponds to the range between $10^{19}$ and $10^{21}$ $m^{-3}$ which are typical for most streamer simulations. The line diagram of the electric field distribution in figure 6 also highlights the electric fields corresponding to the electron density profiles in figure 5. It can be observed that at stable propagation, the electric field in the head of the streamer ranges between 120 and 150 kV/cm. It also highlights the increase in the magnitude of the electric field as the streamer approaches the cathode. This reflects the interaction of the streamer channel with the grounded plane.

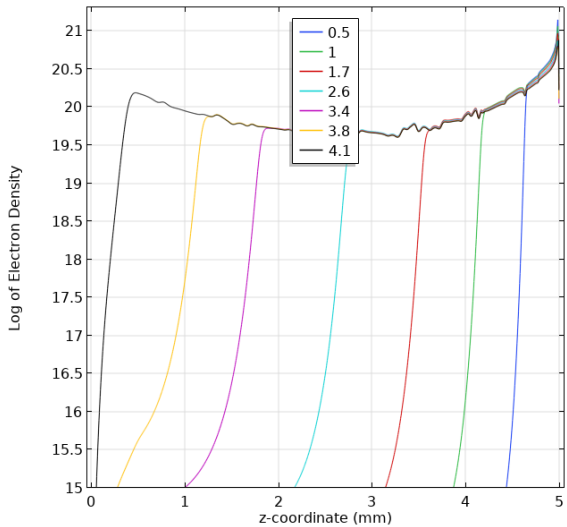

*Figure 5: Line graph of the log of the Electron Density Distribution along the symmetry axis for different time steps.*

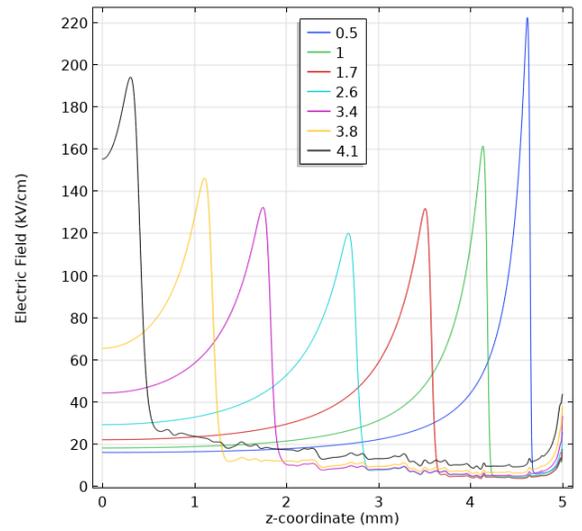

*Figure 6: Line graph of Electric Field Distribution along the symmetry axis corresponding to figure 5.*

### C. Influence of applied voltage

Most of the swarm parameters used in streamer computations are reduced electric field $E/N$ dependent and since the electric field is dependent on the applied voltage (Poisson's Equation), it is of interest to study the influence of the voltage. Four additional voltage levels +11 kV, +13.5 kV, +17.5 kV and +20 kV in addition to the +15 kV in the preceding section are studied for a point to plane geometry separated by a 5 mm channel gap.

Figure 7 shows the time it takes for the streamer head to reach 1 mm, 2 mm, 2.7 mm, 4 mm and to fully bridge the gap. Expectedly as the voltage increases, the time the streamer takes to bridge the gap also reduces. As alluded to earlier, most of the swarm parameters are dependent on the electric field and a higher applied voltage increases the electron drift velocity and effective ionization and consequently the streamer propagation time reduces.

Figure 8 illustrates the evolution of the streamer diameter and velocity in comparison with the quantitative experimental formula for streamer velocity characterization in air $v = 0.5d^2$ $(\frac{mm}{ns})$ where $d$ is the diameter of the streamer [15], [16]. As can be seen, there is conformity between the numerical and empirical results barring a margin of error of about 10%. For example, using this expression, the streamer velocity for the 20 kV applied voltage with a radius of 1 mm should have a velocity of $2\ mm/ns$ and so our computed maximum velocity of $2.27\ mm/ns$ is in good agreement with experimental expectation. The variation can be attributed to simulations measuring the geometrical radius of the space charge while experiments measure the radiative or visible radius.

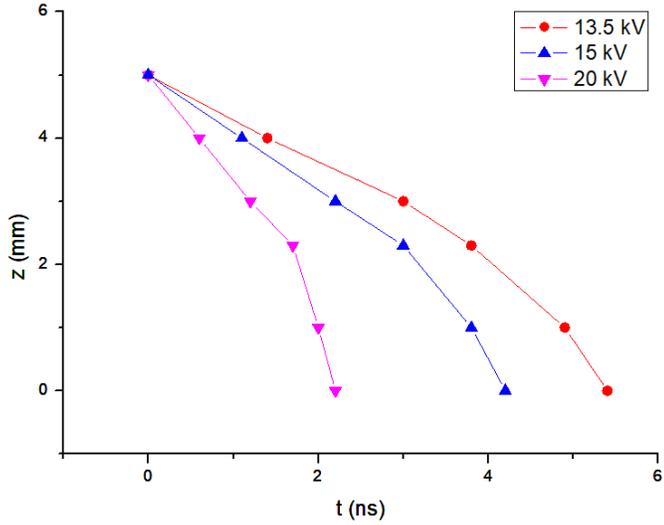

*Figure 7: Position of streamer head as a function of time for applied voltages, 13.5 kV, 15 kV and 20 kV in a 5 mm gap.*

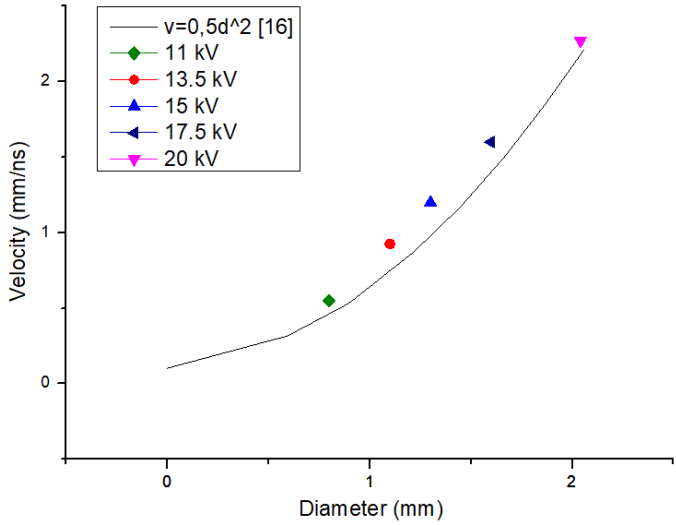

*Figure 8: Plot of streamer velocity versus diameter for varying voltages against empirical fit [16].*

## 5. CONCLUSION

Simulations for streamer discharges are essential for electrical equipment design optimization as they aid in breakdown studies of gases. Established through a huge list of past works, these computations made up of the transport phenomena of charged species coupled with electrostatic computations can be solved efficiently with mathematical simplifications using finite element or finite volume approaches. This has been done in COMSOL™ Multiphysics using the plasma fluid model approach and comparable results to a homemade code in terms of streamer velocity, radius, electric field and charge density in the streamer head have been achieved. Streamers in short gaps have been studied and the influence of applied voltage on the streamer have been expounded on. In concurrence with literature, streamer velocity of $1\,mm/ns$ with a radius of between $400 - 600\,\mu m$ depending on propagation stage was realized. It has been established that, the radius and velocity exhibit direct proportionality to the voltage and this compared to experimental expectations are in agreement.

## 6. ACKNOWLEDGEMENT

This work has been supported by IDEX Université Grenoble Alpes under the Strategic Research Initiatives (IRS) project.

## 7. APPENDIX

### a. Cross Sectional Species

| Reaction Type | Cross Sectional Species |
|---|---|
| Electron Impact Ionization | $e^- + A \to A^+ + 2e^-$ |
| Electron Attachment | $e^- + A \to A^-$ |
| Electron - Ion Recombination | $e^- + A^+ \to A$ |
| Ion - Ion Recombination | $A^- + A^+ \to A + A$ |

### b. Swarm Parameters

- Diffusion $D_e\ (cm^2/s)$ [17]
$$(0.3341 \cdot 10^9 (|E|/N)^{0.54069}) * we/E$$

- Electron Drift velocity $\mu_e E$ (cm/s) [4]
$$-(|E|/E) * (10^{5.5236702 + 0.7822439 \cdot log10(|E|/N)})$$
for $9.8\,Td \leq \frac{E}{N} \leq 1000\,Td$

$$-(|E|/E) * (10^{5.8692884 + 0.4375671 \cdot log10(|E|/N)})$$
for $\frac{E}{N} < 9.8\,Td$

- Reduced Electron Impact Ionization $\frac{\alpha}{N}\ (cm^2)$ [17]
$$2 * 10^{-16} \cdot \exp\left(\frac{-7.248 * 10^{-15}}{|E|/N}\right)$$
for $150\,Td \leq \frac{E}{N}$

$$6.619 * 10^{-17} \cdot \exp\left(\frac{-5.593 * 10^{-15}}{|E|/N}\right)$$
for $\frac{E}{N} < 150\,Td$

- Reduced Electron Attachment $\frac{\eta}{N}\ (cm^2)$ [4]
$$6.56041 \cdot 10^{-19} - 1.45181 \cdot 10^{-21}(E/N) + 1.45951 \cdot 10^{-24}(E/N)^2 - 5.69565 \cdot 10^{-28}(E/N)^3$$
for $600\,Td \leq \frac{E}{N} < 1000\,Td$

$$6.23261 \cdot 10^{-19} - 1.17646 \cdot 10^{-21}(E/N) + 7.51103 \cdot 10^{-25}(E/N)^2$$
for $170\,Td \leq \frac{E}{N} < 600\,Td$

$$-3.611 \cdot 10^{-19} + 1.01192 \cdot 10^{-20}(E/N) - 3.17875 \cdot 10^{-23}(E/N)^2$$
for $69\,Td \leq \frac{E}{N} < 170\,Td$

$$3.10976 \cdot 10^{-19} - 9.41213 \cdot 10^{-21}(E/N) + 1.09693 \cdot 10^{-22}(E/N)^2$$
for $23\,Td \leq \frac{E}{N} < 69\,Td$

$$1.2409 \cdot 10^{-19} + 8.9497$$
$$\cdot 10^{-18} exp(-|E|/N/1.0931) + 1.3216 \cdot 10^{-18}(-|E|/N/6.05148)^2$$

for $1\ Td \leq \frac{E}{N} < 23\ Td$

- Electron-Ion Recombination $\beta_{ep}$ $(m^3/s)$ [17]

$$2e - 13$$

- Ion-Ion Recombination $\beta_{pn}$ $(m^3/s)$ [17]

$$2e - 13$$

- Positive Ion Mobility $\mu_p$ $(\frac{m^2}{Vs})$ [5]

$$2e - 4$$

- Negative Ion Mobility $\mu_n$ $(\frac{m^2}{Vs})$ [5]

$$2e - 4$$

## 8. BIBLIOGRAPHY

[1] G. E. Georghiou, R. Morrow and A. C. Metaxas, A two-dimensional finite element flux corrected transport algorithm for the solution of gas discharge problems, J. Phys. D, vol. 33, no. 19, pp. 2453–2466, 2000.

[2] W. G. Min, H. S. Kim, S. H. Lee and S. Y. Hahn, A study on the streamer simulation using adaptive mesh generation and FEM-FCT, IEEE Trans. Magn., vol. 37, no. 5, pp. 3141–3144, 2001.

[3] U. Ebert, C. Montjin, T. M. Briels, W. Hundsdorfer, B. Meulenbroek, A. Rocco and E. M. van Veldhuizen, The multiscale nature of streamers, Plasma Sources Science and Technology, 15(2), S118., 2006.

[4] O. Ducasse, L. Papageorghiou, O. Eichwald, N. Spyrou and M. Yousfi, Critical analysis on two-dimensional point-to-plane streamer simulations using the finite element and finite volume methods, IEEE Transactions on Plasma Science, 35(5), 1287-1300, 2007.

[5] Y. Serdyuk, Propagation of cathode-directed streamer discharges in air, Proc. COMSOL Conf., Rotterdam, The Netherlands, 2013.

[6] S. Singh, Computational framework for studying charge transport in high-voltage gas-insulated systems, Chalmers University of Technology, 2017.

[7] S. Pancheshnyi, Role of electronegative gas admixtures in streamer start, propagation and branching phenomena., Plasma Sources Science and Technology, 14(4), 2005.

[8] J. Capeillere, P. Ségur, A. Bourdon, S. Célestin and S. Pancheshnyi, The finite volume method solution of the radiative transfer equation for photon transport in non-thermal gas discharges: application to the calculation of photoionization in streamer discharges., Journal of Physics D: Applied Physics, 41(23), 234018., 2008.

[9] O. Ducasse, O. Eichwald and M. Yousfi, Finite volume method for streamer and gas dynamics modelling in air discharges at atmospheric pressure., Finite Volume Method: Powerful Means of Engineering Design, 2012.

[10] J. Fořt, J. Karel, D. Trdlička, F. Benkhaldoun, I. Kissami, J. B. Montavon, K. Hassouni and J. Z. Mezei, Finite volume methods for numerical simulation of the discharge motion described by different physical models., Advances in Computational Mathematics, 45(4), 2163-2189., 2019.

[11] M. Quast and N. R. Lalic, Streamer propagation in a point-to-plane geometry, Excerpt from the Proceedings of the COMSOL Conference, 2009.

[12] Comsol, AB, COMSOL multiphysics user's guide, 2018.

[13] C. M. v. 5.4, Plasma Module User Guide, 2018.

[14] G. V. Naidis, Positive and negative streamers in air: velocity-diameter relation., Physical Review E, 79(5), 057401, 2009.

[15] A. Luque, V. Ratushnaya and U. Ebert, Positive and negative streamers in air: modelling evolution and velocities, J. Phys. D: Appl. Phys. 41, 2008.

[16] T. M. P. Briels, J. Kos, G. J. J. Winands, E. M. van Veldhuizen and U. Ebert, Positive and negative streamers in ambient air: measuring diameter, velocity and dissipated energy., Journal of Physics D: Applied Physics, 41(23), 234004, 2008.

[17] R. Morrow and J. J. Lowke, Streamer propagation in air., Journal of Physics D: Applied Physics, 30(4), 614., 1997.